\documentclass[11pt]{article}
\usepackage{vietnam,epsfig,graphicx}

\def\be{\begin{equation}}
\def\ee{\end{equation}}
\def\bea{\begin{eqnarray}}
\def\eea{\end{eqnarray}}

\begin{document}
% \vspace*{4cm}
\title{STATUS OF THE CKM MATRIX}
%\footnote{Presented at 5th Rencontres du
%Vietnam, Hanoi, August 6--11, 2004. Enrico Fermi Institute Report No.\
%EFI 04-37, hep-ph/0410281.  The present written version updates some results
%presented at the Conference.}}

\author{ JONATHAN L. ROSNER }

\address{Enrico Fermi Institute and Department of Physics,\\
University of Chicago, Chicago, Illinois 60615, USA}

\maketitle\abstracts{The experimental status and theoretical uncertainties
of the Cabibbo--Kobayashi--Maskawa (CKM) matrix describing the charge-changing
weak transitions between quarks with charges $-1/3$ ($d,~s,~b$) and 2/3
($u,~c,~t$) are reviewed.  Some recent methods of obtaining phases of CKM
elements are described.}

\section{Introduction}

Information about the Cabibbo--Kobayashi--Maskawa (CKM) matrix describing the
charge-changing weak transitions between quarks with charges $-1/3$ ($d,~s,~b$)
and 2/3 ($u,~c,~t$) has been steadily improving over the years.  Despite a
wealth of overconstraining experiments, no significant inconsistencies in its
parameters have emerged so far.  One seeks greater accuracy in the
determination of CKM elements not only to expose such inconsistencies, which
could signal new physics, but also to provide input for an eventual theory of
these elements.

The matrix may be defined in one parametrization \cite{Wolfenstein:1983yz} as

\be
V_{\rm CKM} = \left[ \begin{array}{c c c}
V_{ud} & V_{us} & V_{ub} \\
V_{cd} & V_{cs} & V_{cb} \\
V_{td} & V_{ts} & V_{tb} \end{array} \right]
\simeq \left[ \begin{array}{c c c}
1 - \frac{\lambda^2}{2} & \lambda & A \lambda^3 (\rho - i \eta) \\
- \lambda & 1 - \frac{\lambda^2}{2} & A \lambda^2 \\
A \lambda^3 (1 - \bar \rho - i \bar \eta) & - A \lambda^2 & 1 \end{array}
\right]~~~.
\ee
Here $\bar \rho \equiv \rho(1 - \lambda^2/2),~\bar \eta \equiv \eta(1 -
\lambda^2/2)$, with $\lambda \simeq 0.225$, $A \simeq 0.8$, $\bar \eta \simeq
0.36$, $\bar \rho \simeq 0.19$.  Many detailed reviews exist
\cite{revs,HFAG,PDG}; we concentrate on procedures and open questions.

\begin{figure}
\begin{center}
\includegraphics[height=1.7in]{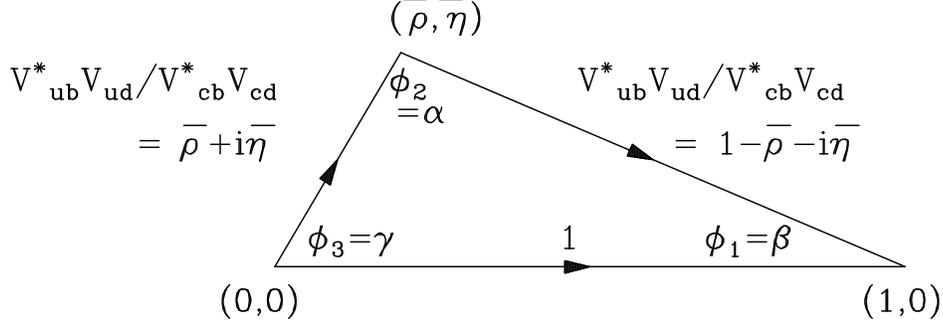}
\end{center}
\caption{Definition of sides and angles of the unitarity triangle.
\label{fig:ut}}
\end{figure}

We shall be concerned with information regarding both magnitudes and
phases of CKM elements.  These are encoded in the angles of
the {\it unitarity triangle}, illustrated in Fig.\ \ref{fig:ut}.
Current fits (not including some CP asymmetries providing information
on $\alpha$ and $\gamma$) imply $1 \sigma$ limits \cite{CKMf}
\be
\beta = (23.8^{+2.1}_{-2.0})^\circ~,~~
\alpha = (94^{+12}_{-10})^\circ~,~~
\gamma = (62^{+10}_{-12})^\circ~~.
\ee
Thus, although $\beta$ is well known [with the ICHEP 2004 average
now $(23.3^{+1.6}_{-1.5})^\circ$] \cite{Pivk}, $\alpha$ and $\gamma$ are
more uncertain, ranging over about $40^\circ$ at the 95\% confidence level
(c.l.).  They can be pinned down more precisely using $B$--$\overline{B}$
mixing, kaon decays, and CP asymmetries in $B$ decays.

\section{$V_{ud}$ from nuclear, neutron, pion $\beta$ decays}

% This is Figure 1

Our discussion is based on a recent overview \cite{Czarnecki:2004cw}.
Nine measurements of nuclear $0^+ \to 0^+$ transitions yield an average of
$|V_{ud}| = 0.9740(1)(3)(4)$, where the errors correspond to experiment,
nuclear theory, and radiative corrections, respectively.  Neutron decay gives
$|V_{ud}|^2(1 + 3 g_A^2) \tau_n = (4908 \pm 4)$ s, so using the measured
lifetime $\tau_n = 885.7(7)$ s and $g_A = 1.2720(18)$ one finds $|V_{ud}| =
0.9729(4)(11)(4)$, where the errors are associated with $\tau_n$, $g_A$, and
radiative corrections.  (A very new value \cite{Serebrov:2004} $\tau_n =
878.5(7)(3)$ s implies $|V_{ud}| = 0.9757(4)(11)(4)$.)  Pion beta decay
($\pi^+ \to \pi^0 e^+ \nu_e$) yields $|V_{ud}| = 0.9739(39)$; an ongoing
experiment at PSI \cite{Frlez:2003vg} seeks to reduce the errors further.
The overall average (before including the result of Ref.\ \cite{Serebrov:2004})
is $|V_{ud}| = 0.9740(5)$.

\section{$V_{us}$: Hyperon and $K_{\ell 3}$ decays; lattice}

Semileptonic hyperon decays, including new measurements of $\Lambda \to p e^-
\bar \nu$, $\Sigma^- \to n e^- \bar \nu$, $\Xi^- \to \Lambda e^- \bar \nu$,
and $\Xi^0 \to \Sigma^+ e^- \bar \nu$, have been analyzed
\cite{Cabibbo:2003ea}, with the result $|V_{us}| = 0.2250 \pm 0.0027$.
Satisfactory fits to all data have been found without the need for SU(3)
breaking, though another analysis \cite{Flores-Mendieta:2004sk} requires it,
obtaining $|V_{us}| = 0.2199 \pm 0.0026$.  One remaining question is the
magnitude of the axial
weak-magnetism parameter $g_2$, which is not well constrained by data.

Several experiments have remeasured $K_{\ell 3}$ decays.  Brookhaven
E865 \cite{Sher:2003fb} finds $|V_{us}| = 0.2272 \pm 0.0022_{\rm rate}
\pm 0.0007_{\rm f.f.} \pm 0.0018_{f_+(0)}$;
radiative corrections were an important part of the analysis.  Fermilab
E832 (the KTeV Collaboration) \cite{Alexopoulos:2004sw}
obtains $|V_{us}| = 0.2252 \pm 0.0008_{\rm KTeV} \pm
0.0021_{\rm ext}$, where ``ext'' refers to all errors external to those in
KTeV, such as uncertainties in form factors, making use of the
radiative corrections in \cite{Andre:2004tk}. 

The value of $|V_{ud}| = 0.9740 \pm 0.0005$, when combined with the Particle
Data Group \cite{PDG} value $|V_{us}| = 0.2200 \pm 0.0026$ agreed poorly with
CKM unitarity: $|V_{ud}|^2 + |V_{us}|^2 + |V_{ub}|^2 = 0.9971 \pm 0.0015$.
while the expected value of $|V_{us}|$ from unitarity is $|V_{us}| = (1 -
|V_{ud}|^2 - |V_{ub}|^2)^{1/2} = 0.2265 \pm 0.0023$.  The above two
$K_{\ell 3}$ results, as well as that \cite{Cabibbo:2003ea} from hyperon
decays (see, however, Ref.\ \cite{Flores-Mendieta:2004sk}), are
much more consistent with unitarity.  So, too, are very recently reported
results from the KLOE detector at the DA$\Phi$NE $e^+ e^-$ collider at Frascati
\cite{Patera}.  The NA48 Collaboration \cite{Lai:2004px}
reports a value $|V_{us}| = 0.2187 \pm 0.0028$ using a value of $f_+(0)$ about
1.35\% higher than that used by KTeV, so the situation is still not entirely
settled.

\section{$V_{cd}$ and $V_{cs}$: Charm and $W$ decays; neutrino production}

The CLEO III detector has reported rates and spectra for $D^0 \to \pi^- \ell^+
\nu$ and $D^0 \to K^- \ell^+ \nu$ \cite{Huang:2004fr} implying
$|f^\pi_+(0)|^2 |V_{cd}|^2/|f^K_+(0)|^2 |V_{cs}|^2
= 0.038^{+0.006+0.005}_{-0.007-0.003}$.  Progress in $n_f = 3$ lattice QCD by
the Fermilab-MILC Collaboration \cite{Okamoto} has yielded form factors
$f_+^{D \to \pi}(0) = 0.64(3)(5)$, $f_+^{D \to K}(0) = 0.73(3)(6)$, where the
lattice errors are statistical and systematic, leading when combined with
previous measurements of nonstrange and strange charm decays to
$|V_{cd}| = 0.239(10)(19)(20)$, $|V_{cs}| = 0.969(39)(78)(24)$, where the
final errors are experimental.

Charm production by neutrinos (signaled by dileptons) leads \cite{PDG} to
$\overline{\cal B}(c \to \ell + X) |V_{cd}|^2 = (4.63 \pm 0.34) \times 10^{-3}$
so that with $\overline{\cal B}(c \to \ell + X) = (9.23 \pm 0.73)\%$ one has
$|V_{cd}| = 0.224 \pm 0.012$.  New measurements by the CLEO-c Collaboration of
${\cal B}(D^0 \to [K^-,\pi^-] e^+ \nu)$ and ${\cal B}(D^+\to [K^{*0},\rho^0]
e^+ \nu)$ with 57 pb$^{-1}$ at the $\psi''(3770)$ \cite{Shipsey}
represent a further source of information on $V_{cd}$ and $V_{cs}$ when
combined with the lattice results \cite{Dytman}.  For example, the CLEO-c
result for ${\cal B}(D^0 \to \pi e \nu)/{\cal B}(D^0 \to K e \nu)$ is $0.070
\pm 0.007 \pm 0.003$, to be compared with CLEO-III's $0.082 \pm 0.006 \pm
0.005$.  Higher-luminosity running in $e^+ e^- \to \psi''(3770) \to D \bar D$
is to begin in September, and further improvements are envisioned.  The CLEO-c
Collaboration eventually hopes for 3 fb$^{-1}$ at the $\psi''$.  Lattice errors
will be the limiting factor in extracting $|V_{cd}|$ to ${\cal O}(1\%)$.

Charm-tagged $W$ decays at LEPII (ALEPH,DELPHI) \cite{PDG} have given the value
$|V_{cs}| = 0.97 \pm 0.09 \pm 0.07$.  Measurement of the leptonic branching
ratio ${\cal B}(W \to \ell \nu)$ and the assumption of a standard
pattern of $W$ decays can be used to improve this estimate through the relation
\be
\frac{1}{{\cal B}(W \to \ell \nu)} = 3 \left(1 + \left[ 1 +
\frac{\alpha_s(M_W)}{\pi} \right]
\sum |V_{ij}|^2 \right)~~~(i=u,c;~j=d,s,b)
\ee
which implies $\sum |V_{ij}|^2 = 2.039 \pm 0.025$ and hence $|V_{cs}| = 0.996
\pm 0.013$ when contributions of other CKM elements are subtracted.  The study
of Cabibbo-favored ($c \to s$) decays of charmed particles at CLEO-c will
provide $|V_{cs}|$ at ${\cal O}(1\%)$ accuracy if the ${\cal L}$ goal is
achieved and if lattice gauge theory continues to progress.

At this juncture one can't see a violation of two-family unitarity:
\be
|V_{ud}|^2 + |V_{us}|^2 = |V_{cd}|^2 + |V_{cs}|^2 =
|V_{ud}|^2 + |V_{cd}|^2 = |V_{us}|^2 + |V_{cs}|^2 =1~.
\ee
This reflects on the very hierarchical structure of the CKM matrix.  It implies
that violations of unitarity may be too small to signal the presence of
additional quark families.

\section{$V_{cb}$: $b \to c$ inclusive and exclusive decays}

At the quark level, the $b \to c$ semileptonic decay rate is
simple (neglecting $m_\ell$):
\be
\Gamma (b \to c \bar \nu_\ell \ell^-) =
\frac{G_{F}^2 m_{b}^5}{192 \pi^3} | V_{cb} |^2 f \left( \frac{m^2_c}{m^2_b}
\right ) ~~,~~~
f(x) \equiv 1 - 8 x + 8 x^3 - x^4 - 12 x^2 \ln x~~.
\ee
However, initial and final states contain real hadrons, so one must employ
perturbative QCD; expansions in inverse heavy quark mass (``HQE''); and
moments in lepton energy, hadron mass, and photon energy in $b \to s \gamma$
(see, e.g., \cite{Ali04} for a review of these techniques) to infer the 
inclusive value $|V_{cb}| = 0.0421 \pm 0.0013$ \cite{KS}, which is quoted in
\cite{CKMf} as $0.0420 \pm 0.0006_{\rm stat} \pm 0.0008_{\rm theo}$.  More
recent contributions include $|V_{cb}| = (41.4 \pm 0.4_{\rm stat} \pm 0.4_{\rm
HQE} \pm 0.6_{\rm theo}) \times 10^{-3}$ \cite{Flaecher:2004sw} and $|V_{cb}|
=(42.4 \pm 0.8_{\rm stat+HQE} \pm (>0.8)_{\rm theo}) \times 10^{-3}$
\cite{CLEOVcb}.

The best source of $|V_{cb}|$ from exclusive decays is the process $B \to D^*
\ell \bar \nu_\ell$, whose rate is
\be
\frac{d\Gamma}{dw} = \frac{G^2_F}{4 \pi^3}|V_{cb}|^2 (m_B - m_D^*)^2 m_{D^*}^3
\sqrt{w^2 - 1} {\cal G}(w) |{\cal F}(w)|^2~~.
\ee
One Isgur-Wise form factor ${\cal F}(w) \simeq {\cal F}(1) [1 + \rho^2(w-1)
\ldots]$, a function of the variable $w = v_B \cdot v_{D^*}$, governs the
process; phase space is described by a function ${\cal G}(w)$ with ${\cal G}(1)
= 1$.  The form factor at $w=1$ is given by ${\cal F}(1) = \eta_{QCD}[1 + {\cal
O}(1/m_b^2)] = 0.913^{+0.030}_{-0.035}$ as calculated using lattice and HQET
estimates \cite{Hashimoto:2001nb}.

\begin{figure}
\begin{center}
\includegraphics[height=4in]{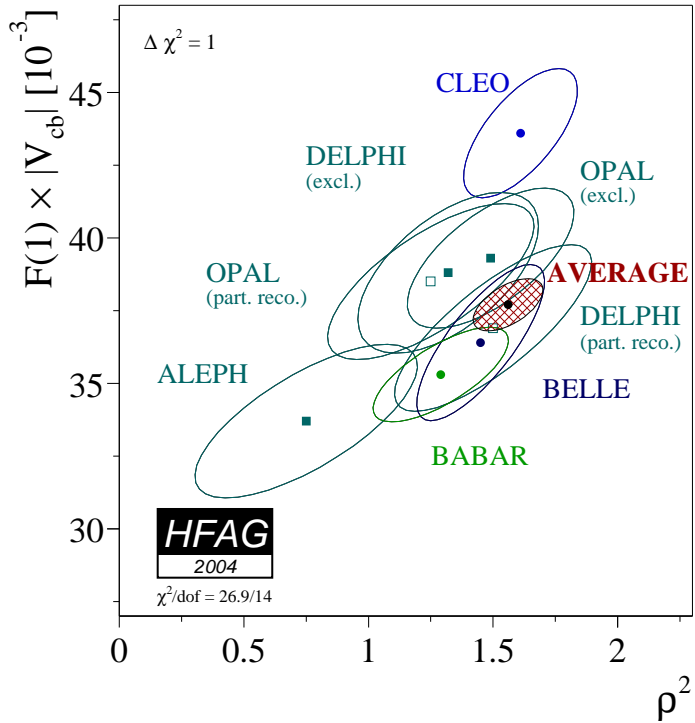}
\end{center}
\caption{Values of ${\cal F}(1) |V_{cb}|$ vs.\ form factor parameter $\rho^2$
measured by various groups.
\label{fig:Vcb}}
\end{figure}

A compilation of values of ${\cal F}(1) |V_{cb}|$ \cite{HFAG} is shown in Fig.\
\ref{fig:Vcb}.
% This is Figure 2
The latest average is ${\cal F}(1) |V_{cb}|^2=(37.7 \pm 0.9) \times 10^{-3}$,
with the form factor slope parameter $\rho^2 = 1.56 \pm 0.14$, leading to
$|V_{cb}| = (41.4 \pm 1.0_{\rm exp} \pm 1.8_{\rm th}) \times 10^{-3}$.  This
form factor shape is consistent with the rates for $B^0 \to D^{(*)-}
\pi^+,~D^{(*)-} D_s^{(*)+}$ estimated with a factorization approach
\cite{Luo:2001mc}.  It is consistent with the inclusive $|V_{cb}|$ value.
Although it currently has larger errors, progress in experiment and lattice
estimates may eventually make this the best source of $|V_{cb}|$.

\section{$V_{ub}$: $B \to u$ inclusive and exclusive decays}

The semileptonic decay $b \to u \bar \nu_\ell \ell^-$ is the source of
information on $V_{ub}$ from inclusive decays.  However, $\Gamma(b \to u \bar
\nu_\ell \ell^-)$ is only about 2\% of $\Gamma(b \to c \bar \nu_\ell \ell^-)$.
Several strategies have been used to extract the $b \to u \bar \nu_\ell \ell^-$
contribution \cite{Ali04}, including measurement of leptons beyond the $b \to c
\bar \nu_\ell
\ell^-$ end point, reconstruction of hadronic masses $M_X < M_D$, cutting on
$q^2 = m^2_{\ell \nu}$, and, most recently, light-cone-inspired kinematic
cuts \cite{Bosch:2004th}.  As in $b \to c \bar \nu_\ell \ell^-$, the $b \to s
\gamma$ photon spectrum helps to pin down hadronic uncertainties.

\begin{figure}
\begin{center}
\includegraphics[height=4.7in]{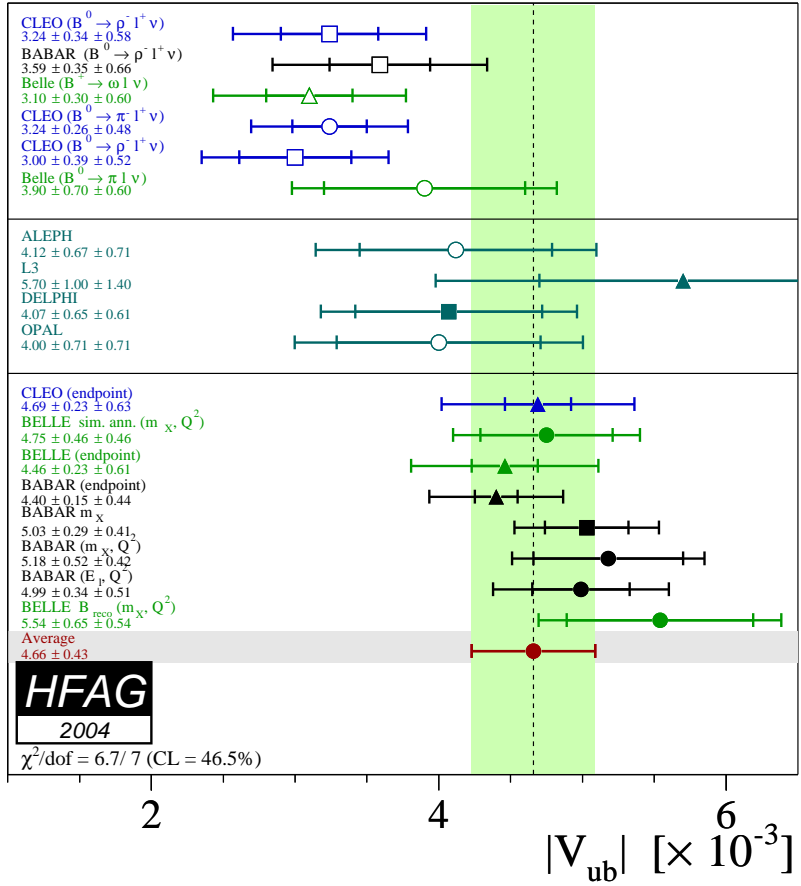}
\end{center}
\caption{Values of $|V_{ub}|$ measured by various groups.
\label{fig:Vub}}
\end{figure}

A compilation of values of $|V_{ub}|$ \cite{HFAG} is shown in Fig.\
\ref{fig:Vub}.  The bottom 5 points lead to an average $|V_{ub}| = (4.66 \pm
0.43) \times 10^{-3}$.  This is notably higher than the value from exclusive
$b \to u$ decays (e.g., $B \to \pi \bar \nu_\ell \ell$).  The top 6 plotted
points give an average of $|V_{ub}| = (3.26 \pm 0.62) \times 10^{-3}$ from
exclusive decays.  In this average I assigned an overall systematic error of
$\pm 0.60$.  Combining with the inclusive value and including a scale factor
\cite{PDG} $S=1.86 = \sqrt{\chi^2}$, I find $|V_{ub}| = (4.21 \pm 0.66)
\times 10^{-3}$.  The slight discrepancy between inclusive and exclusive values
merits caution; quark-hadron duality may not be valid if extended down to
hadronic masses which are so low as to be represented by discrete states like
$\pi$ and $\rho$.

The measurement of the spectrum for $B^0 \to \pi^- l^+ \nu_l$, e.g., as
performed by CLEO \cite{Athar:2003yg}, can be useful not only in extracting
$|V_{ub}|$ from lattice gauge theory form factor results (typically obtained
for high $q^2$), but also in testing factorization in $B^0 \to \pi^+ \pi^-$
\cite{Luo:2003hn}.

\section{$V_{td}$: $B^0$--$\overline{B}^0$ mixing; progress on decay constants}

Loop diagrams with quarks $i,j = u,c,t$ in the intermediate state allow $b \bar
d \leftrightarrow d \bar b$ transitions at 2nd order in weak interactions.
The $t$ quark dominates, so this mixing provides information on $|V_{td}|$.
The predicted splitting $\Delta m_d$ between mass eigenstates in the
$B^0$--$\overline{B}^0$ system is proportional to a parameter $f^2_B$ (the $B$
meson decay constant) governing the matrix element of the $b \bar d
\leftrightarrow d \bar b$ operator between physical meson states, and to a
parameter $B_B$ equal to 1 if $W$ exchange diagrams dominate.
A lattice estimate $f_B \sqrt{B_B} = (228 \pm 30 \pm 10)$ MeV and the
experimental value $\Delta m_d \simeq 0.5$ ps$^{-1}$ imply the 95\% c.l.\ range
\cite{CKMf} $|V_{td}| = (8.26^{+1.23}_{-1.79}) \times 10^{-3}$, equivalent to
$|1 - \bar \rho - i \bar \eta| = 0.89^{+0.12}_{-0.20}$.  (The upper limit on
these quantities is governed by a lower limit on $B_s$--$\overline{B}_s$
mixing to be discussed below.)  For comparison, the 95\% c.l.\ range
\cite{CKMf} of $|V_{ub}| = (3.87^{+0.73}_{-0.61}) \times 10^{-3}$ implies
$|\bar \rho + i \bar \eta| = 0.40^{+0.08}_{-0.06}$.
Information from ${\cal B}(B \to \rho \gamma)/ {\cal B}(B \to K^* \gamma)$
also constrains $|V_{td}/V_{ts}|$.  One expects, for example,\cite{Ali}
${\cal B}(B^0 \to \rho^0 \gamma) = (0.64 \pm 0.23) \times 10^{-6}$.  A recent
report \cite{Aubert:2004fq} finds ${\cal B}(B^0 \to \rho^0 \gamma) < 0.4
\times 10^{-6}$ and $|V_{td}/V_{ts}| < 0.19$ at 90\% c.l.

% This is Figure 3

The new CLEO value $f_D = (201 \pm 41 \pm 17)$ MeV \cite{CLEOfd}
has an accuracy approaching that of lattice calculations.  With increased
integrated CLEO luminosity, this quantity will be measured precisely enough
to test those calculations, lending weight to their predictions for $f_B$.

As mentioned, $|V_{td}|$ is quoted with an asymmetric error since it
is restricted on the positive side by $B_s$--$\overline{B}_s$ mixing.  This
quantity is described by the same diagram as $B^0$--$\overline{B}^0$ mixing
with the substitution $d \to s$.  We assume $V_{ts} \simeq -V_{cb}$.  Then the
lower limit $\Delta m_s > 14.5$ ps$^{-1}$ and the SU(3)-breaking estimate $\xi
\equiv f_{B_s} \sqrt{B_{B_s}}/f_B \sqrt{B_B} = 1.21 \pm 0.06$ implies a
lower limit on $V_{ts}/V_{td}$ and hence an upper limit on $V_{td}$.  The
($\pm 2 \sigma$) prediction of Ref.\ \cite{CKMf} is
$\Delta m_s = 17.8^{+15.2}_{-2.7}$ ps$^{-1}$.

\begin{figure}
\begin{center}
\includegraphics[height=1.6in]{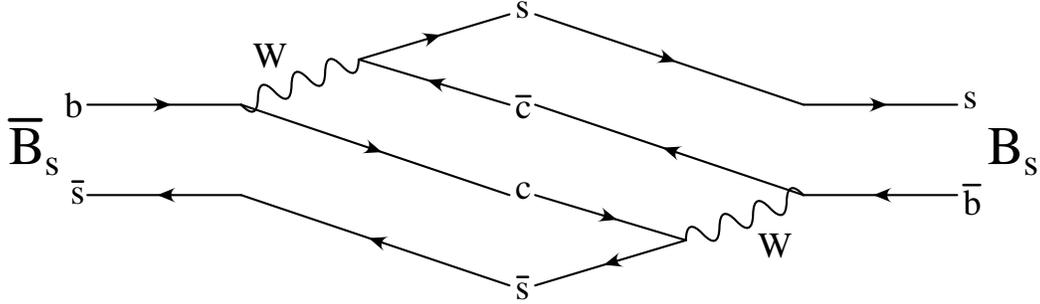}
\end{center}
\caption{One graph describing $B_s$--$\overline{B}_s$ mixing, which can
correspond to a real intermediate $c \bar c s \bar s$ state.
\label{fig:bsmix}}
\end{figure}

The mixing of $B_s$ and $\overline{B}_s$ can proceed via on-shell shared
intermediate states, as described in Fig.\ \ref{fig:bsmix}.  One expects (see,
e.g., \cite{Dighe:1995pd} and references therein) $\Delta \Gamma_s \simeq -
\Delta m_s /200$ ($\sim m_b^2/m_t^2$) or $\Delta \Gamma_s / \bar \Gamma_s\simeq
0.18 (f_{B_s}/200~{\rm MeV})^2$ in lowest order, but a state-of-the-art
calculation \cite{Lenz:2004} finds $\Delta \Gamma_s / \bar \Gamma_s = 0.12 \pm
0.05$ for $f_{B_s} = 245$ MeV.  The CDF Collaboration has recently reported a
value of $0.65^{+0.25}_{-0.23} \pm 0.01$ \cite{Gay}.  Since $f_{B_s}$ is
expected to be no larger than about 300 MeV, the CDF result is somewhat larger
than anticipated, but not yet in serious conflict with theory.

% This is Figure 4

\section{Constraints from $K_L$, $K^+$ decays}

\begin{figure}
\begin{center}
\includegraphics[height=2.5in]{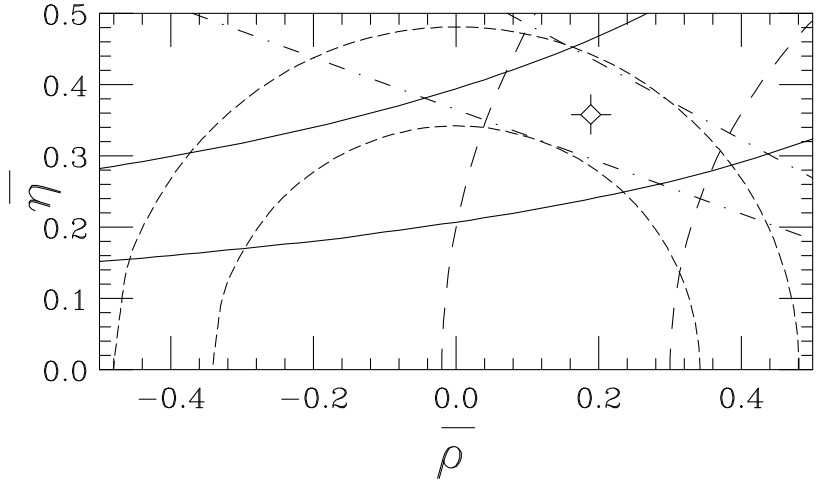}
\end{center}
\caption{Summary of 95\% c.l.\ constraints on $(\bar \rho, \bar \eta)$ due to
$|V_{ub}|$ (short-dashed circles), $B^0$--$\overline{B}^0$ and
$B_s$--$\overline{B}_s$ mixing (dashed circles), CP-violating
$K^0$--$\overline{K}^0$ mixing (solid hyperbolae, $\pm 1 \sigma$ limits),
and $\sin 2 \beta$ from CP asymmetry in $B^0 \to J/\psi K_S$ and
related processes (dash-dotted lines).
\label{fig:constr}}
\end{figure}

CP-violating $K^0$--$\overline{K}^0$ mixing is dominated by top quarks in
second-order-weak loop diagrams.  The parameter $\epsilon_K$ describing this
mixing depends mainly on Im($V_{td}^2$) and hence measures approximately $\bar
\eta (1 - \bar \rho)$, with charmed quarks supplying a small correction.
Convenient expressions in Ref.\ \cite{Ali} imply that $(\bar \rho,~ \bar \eta)$
lies between the $1 \sigma$ boundaries $\bar \eta(1.38 - \bar \rho) = 0.28$ and
$\bar \eta(1.26 - \bar \rho) = 0.50$.  These constraints and those on $(\bar
\rho,~ \bar \eta)$ from $|V_{ub}|$, $|V_{td}|$, and $\sin 2 \beta$ select a
region around $(0.19^{+0.09}_{-0.07},0.36^{+0.05}_{-0.04})$, shown as the
plotted point in Fig.\ \ref{fig:constr}.

% This is Figure 5

Note the consistency of all these determinations.  The main uncertainty is
associated with the value of $|1 - \rho - i \eta|$.  Direct CP violation in
neutral kaon decay (governed by a parameter$\epsilon'/\epsilon$) is seen but
provides no useful constraint as a result of hadronic uncertainties.

In $K \to \pi \nu \bar \nu$ decays, higher-order weak diagrams govern the weak
quark transition $s \to d \nu \bar \nu$.  For $K^+ \to \pi^+ \nu \bar \nu$ the
top quark in the loop dominates but there is also a charm contribution, so that
the rate measures $|1.3 - \bar \rho - i \bar \eta|$.  An experiment (E787,
E949) at Brookhaven sees three events \cite{Anisimovsky:2004hr}, corresponding
to a branching ratio ${\cal B}(K^+ \to \pi^+ \nu \bar \nu) =
(1.47^{+1.30}_{-0.82}) \times 10^{-10}$.  Comparing with the Standard Model
prediction \cite{Buras:2004uu} of $(0.78 \pm 0.12) \times 10^{-10}$,
the result is still consistent with a large $(\bar \rho,~\bar \eta)$ region.
Proposals for Fermilab and JPARC seek a sample of 100 events which
could determine $|1.3 - \bar \rho - i \bar \eta|$ to 5\%.

The process $K_L \to \pi^0 \nu \bar \nu$ is purely CP-violating and
measures $\bar \eta^2$.  One expects ${\cal B} = (3.0 \pm 0.6) \times 10^{-11}$
\cite{Buras:2004uu}.  An experiment at KEK (PS E391) has taken data whose
single-event sensitivity should be an order of magnitude above the Standard
Model value \cite{Patera}; this represents tremendous progress in the past few
years.  The eventual goal of proposed experiments at JPARC and Brookhaven is
to be sensitive at the Standard Model level.

\section{$V_{ts}$ and $V_{tb}$:  $b \to s \gamma$, top quark decays, unitarity}

One may obtain a lower limit on $|V^*_{tb} V_{ts}|$ from $B_s$--$\bar B_s$
mixing, the assumption that $\xi =
1.2 \pm 0.1$, and $B_d$--$\bar B_d$ mixing, yielding \cite{Ali}
$|V^*_{tb} V_{ts}| > 0.034$.  An upper limit can be extracted from the top
quark contribution to $b \to s \gamma$ \cite{AM}:  $V^*_{tb} V_{ts} = -0.047
\pm 0.008$, which has the expected sign.

The fact that top quark decays $t \to b \ell^+ \nu_\ell$ dominate over those
with no $b$ has been used by the CDF Collaboration to conclude on the basis
of Run I data that $|V_{tb}|^2/(|V_{td}|^2 + |V_{ts}|^2 + |V_{tb}|^2) =
0.94^{+0.31}_{-0.24}$.
A slightly less stringent limit has been placed for a sample of 108 pb$^{-1}$
of Run II data:  $0.54^{+0.49}_{-0.39}$.  The assumption that CKM matrix is $3
\times 3$ is crucial to interpretation of these results as a useful bound on
$|V_{tb}|^2$.  The lower bounds on magnitudes of the third row of the CKM
matrix are almost non-existent if there are more than 3 families \cite{PDG}.

\section{Processes giving $\beta$}

Many experiments measure a time-dependent CP asymmetry in $B$ decays of the
form $A(t) = -C \cos(\Delta mt) + S \sin(\Delta mt)$.  Results involving the
subprocess $b \to c \bar c s$, such as $B^0 \to J/\psi K_S$, show beautiful
agreement with Standard Model fits, as illustrated in Fig.\ \ref{fig:constr}.

% This is Figure 6
\begin{figure}
\includegraphics[width = 0.49\textwidth]{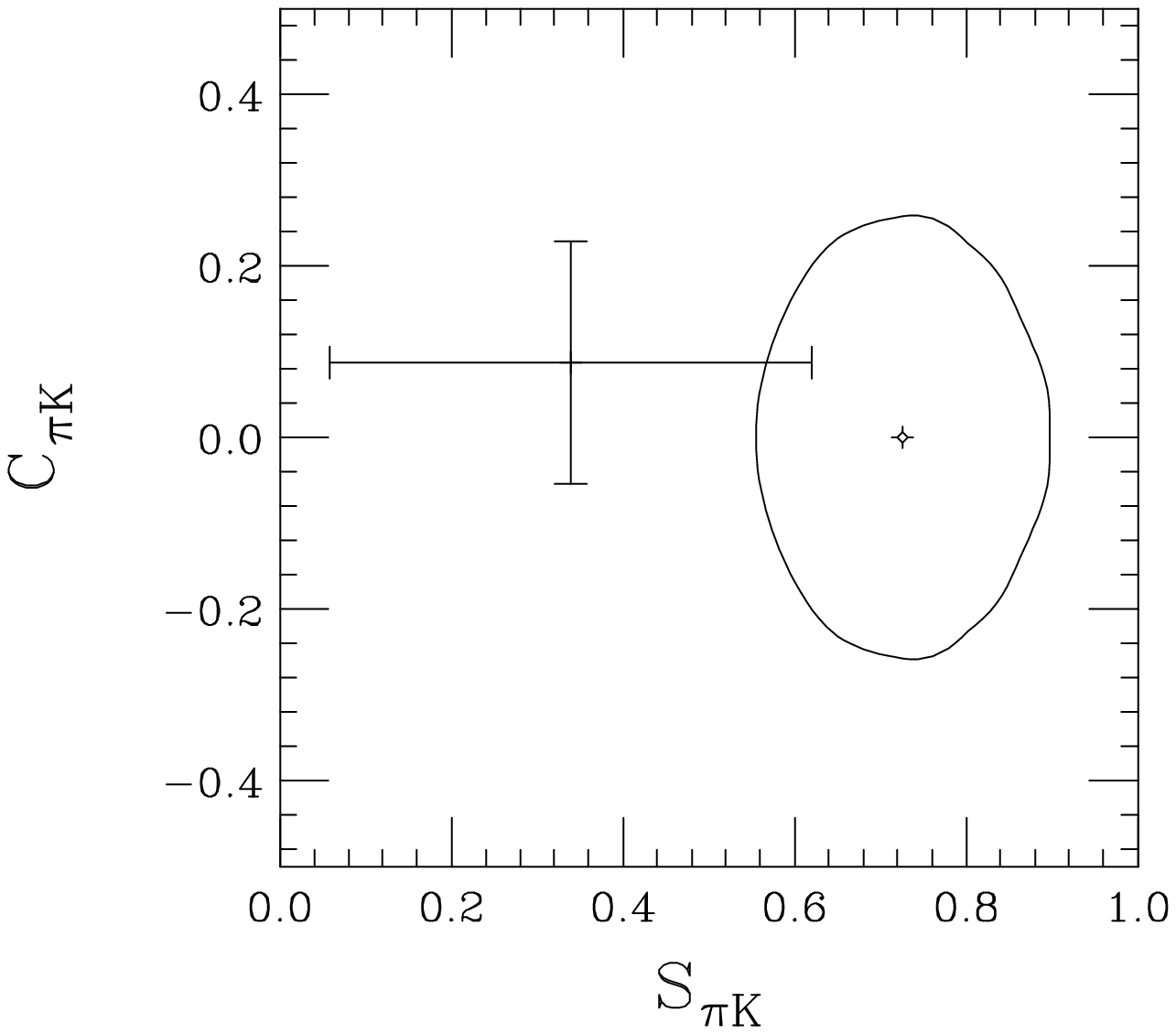}
\includegraphics[width = 0.49\textwidth]{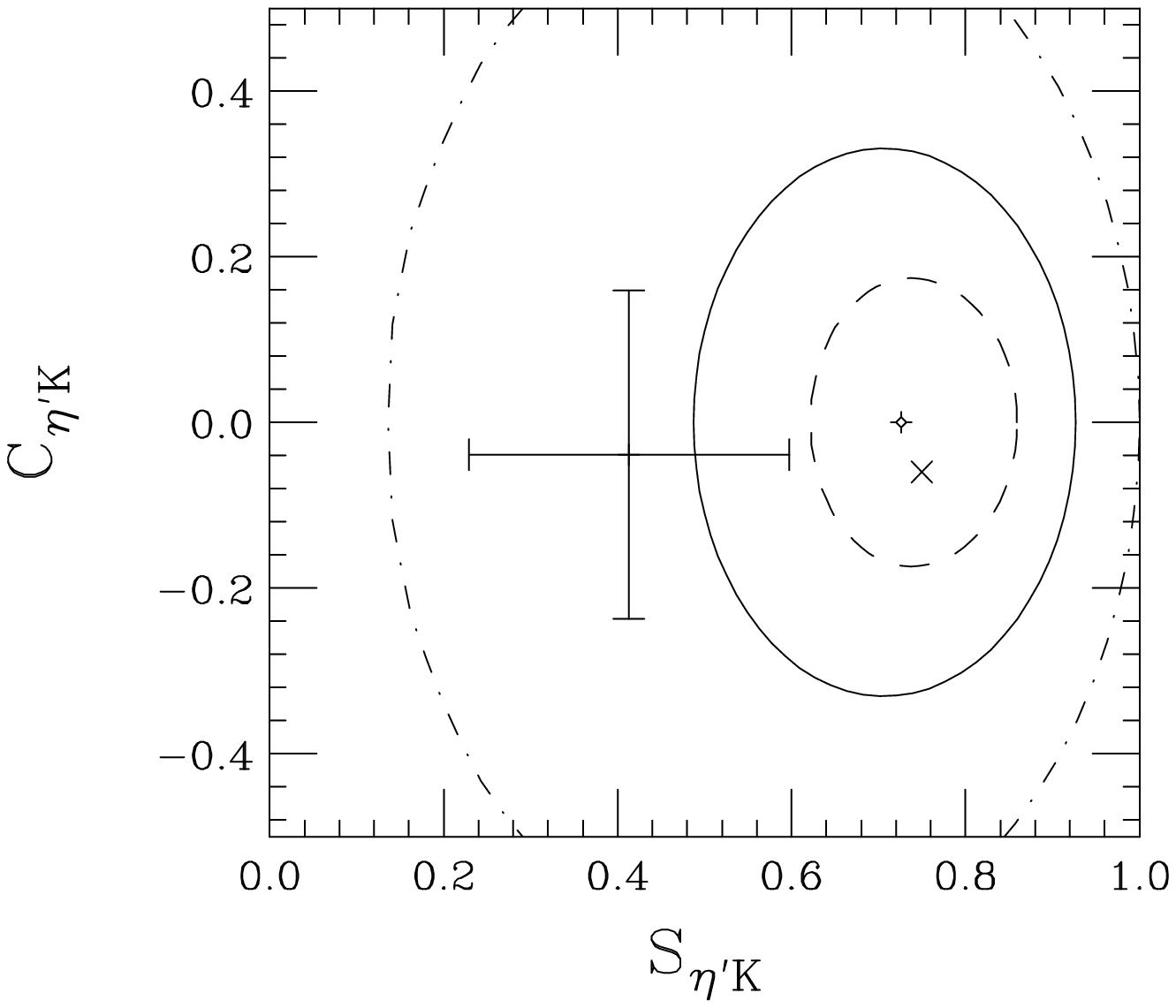}
\caption{Left: $B^0 \to \pi^0 K_S$; solid ellipse: allowed region boundary.
Right: $B^0 \to \eta' K_S$;  solid curve: bounds based on flavor SU(3) and
measurements of other processes; dot-dashed curve:  bounds prior to latest set
of such measurements; dashed curve: neglecting processes involving
spectator quark.
\label{fig:betaprocs}}
\end{figure}

For processes dominated by the $b \to s$ penguin amplitudes one expects $C =
0$, $S = \sin(2 \beta) = 0.74 \pm 0.05$.  One can estimate contributions from
other amplitudes using new measurements and flavor SU(3), as in the processes
$B^0 \to \pi^0 K_S$ \cite{Gronau:2003br} and $B^0 \to \eta' K_S$
\cite{Gronau:2004hp}.  We show in Fig.\ \ref{fig:betaprocs} examples of bounds
on allowed deviations from the Standard Model predictions.  We have applied a
scale factor in averaging $\eta' K_S$ results from BaBar \cite{Giorgi}
($S_{\eta' K_S} = 0.27 \pm 0.14 \pm 0.03$, $C_{\eta' K_S} = -0.21 \pm 0.10 \pm
0.03$) and Belle \cite{Sakai} ($S_{\eta' K_S} = 0.65 \pm 0.18 \pm 0.04$,
$C_{\eta' K_S} = 0.19 \pm 0.11 \pm 0.05$), so the plotted data points have
larger error bars than quoted in, e.g., Ref.\ \cite{HFAG}).  The corresponding
values for $\pi^0 K_S$ are $S_{\pi^0 K_S} = 0.35^{+0.30}_{-0.33}
\pm 0.04$, $C_{\pi^0 K_S} = -0.21 \pm 0.10 \pm 0.03$ (BaBar \cite{Giorgi})
$S_{\pi^0 K_S} = 0.30 \pm 0.59 \pm 0.11$, $C_{\pi^0 K_S} = 0.12 \pm 0.20 \pm
0.07$ (Belle \cite{Sakai}).  There is not, in my opinion, evidence yet for
substantial deviations from the Standard Model predictions, but the situation
bears watching both in the processes illustrated in Fig.\ \ref{fig:betaprocs}
and in the decay $B^0 \to \phi K_S$, also dominated by the $b \to s$ penguin.
With improved data one could well have evidence for new physics.

\section{$B^0 \to \pi^+ \pi^-, \pi^\pm \rho^\mp, \rho^+ \rho^-$ and $\alpha$}

The time-dependent parameters $S$ in the processes $B^0 \to \pi^+ \pi^-$, $B^0
\to \rho^\pm \pi^\mp$, and $B^0 \to \rho^+ \rho^-$ would measure just $\sin 2
\alpha$ if one could neglect the effect of penguin ``pollution'' of the
dominant tree amplitudes.  In $B \to \pi \pi$ the penguin/tree amplitude
ratio was estimated to be $\sim 70\%$ from $B \to K \pi$ decays
\cite{Gronau:2004ej}, leading to the estimate $\alpha = (103 \pm 17)^\circ$.
In $B^0 \to \rho^\pm \pi^\mp$ the relative penguin contribution is found to be
less \cite{Gronau:2004tm}, leading to $\alpha = (95 \pm 16)^\circ$, while in
$B^0 \to \rho^+ \rho^-$ a recent BaBar measurement, combined with an
estimate of penguin effects, gives \cite{Aubert:2004zr} $\alpha = (96 \pm 10
\pm 4 \pm 13)^\circ$.

\section{$B \to PP, VP$ and $\gamma$}

\begin{figure}
\begin{center}
\includegraphics[width=0.39\textwidth]{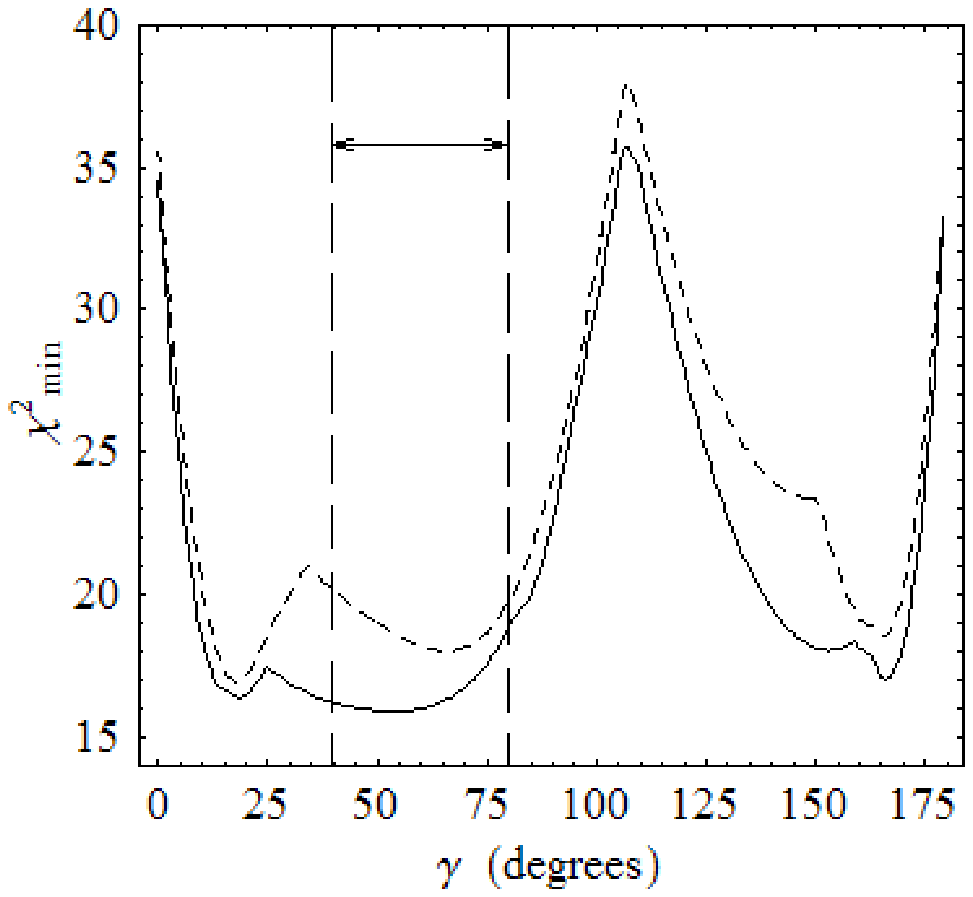}
\includegraphics[width=0.475\textwidth]{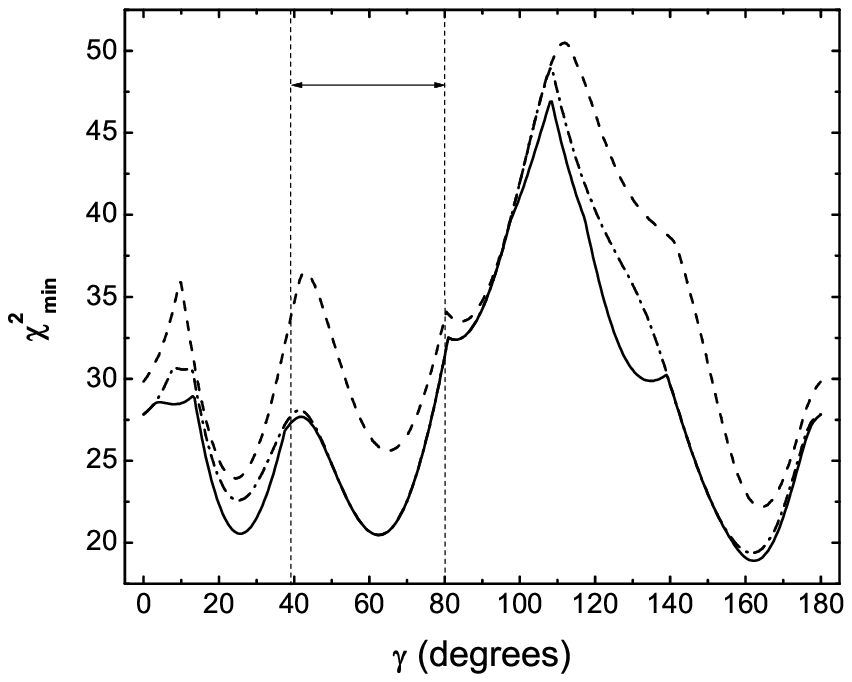}
\end{center}
\caption{Results of $\chi^2$ fits to $B \to PP$ (left) and $B \to VP$ (right)
decays.  In $PP$ fit the lower curve corresponds to a fit with two extra
parameters.  In $VP$
fit the solid curve corresponds to a fit with no constraint between two
independent penguin amplitudes, while the dot-dashed and dashed curves
correspond respectively to constraining these amplitudes to be relatively
real, and equal and opposite.
\label{fig:PPVP}}
\end{figure}

One can perform fits to rates and CP asymmetries in flavor SU(3) based on
amplitudes designated by $T,~P,~C, \ldots$ (denoting tree, penguin,
color-suppressed, etc.), for $B \to PP$ \cite{Chiang:2004nm} and $B \to VP$
\cite{Chiang:2003pm} decays, where $P,V$ denote light (pseudoscalar, vector)
mesons composed of $u,~d,~s$ quarks.  The $B \to PP$ fits involve 26
observables while the $B \to VP$ fits involve 34.  The results are shown in
Fig.\ \ref{fig:PPVP}.  We place greatest reliance on the fits dscribed by
the solid curves, corresponding to $\chi^2_{\rm min}/{\rm d.f.} =
16.0/13$ (left) and $\chi^2_{\rm min}/{\rm d.f.} = 20.5/22$ (right).  The
fit to $PP$ decays gives a rather shallow minimum around $\gamma = 54^\circ$
whose stability we do not yet trust. while the fit to $VP$ decays gives $\gamma
= (63 \pm 6)^\circ$.  The combined fit gives local minima at $\gamma \simeq
55^\circ$ and $61^\circ$ after being updated in view of new data presented at
ICHEP 04 \cite{Suprun}.

% This is Figure 7

\section{Beyond the $3 \times 3$ CKM matrix}

Does a fourth family of quarks and leptons exist?  Any fourth neutrino
must be heavy;  only 3 light neutrinos are seen in $Z$ decay \cite{EW}.
A direct search for $b'$ heavier than $Z$ (looking for $b' \to b Z$) by CDF at
the Tevatron \cite{Affolder:1999bs} excludes $100 < m(b') < 199$ GeV/$c^2$.
However, looking outside the familiar pattern, existing quark--lepton families
belong to 16-dimensional multiplets of the grand unified group SO(10),
consisting of {\bf 1} + {\bf 5$^*$} + {\bf 10} of SU(5).  The smallest
representation {\bf 27} of E$_{\rm 6}$, an interesting group sontaining SO(10),
is {\bf 16} + {\bf 10} + {\bf 1} of SO(10).  The {\bf 10} of SO(10) consists
of isosinglet quarks ``$h,\bar h$'' with charge $Q = \pm 1/3$, and isodoublet
leptons.  The SO(10) singlets are candidates for sterile neutrinos, one
for each family.

The exotic $h$ quarks can mix with $b$ and push its mass down with respect to
$t$ \cite{Rosner:2000rd}.  Production signatures of $h \bar h$ at the
Tevatron and LHC have been investigated \cite{Andre:2003wc}.  Through decays
of $h$ to $Z + b$, $W + t$, and possibly ${\rm Higgs} + b$, one can reach
$h$ masses 270--320 GeV/$c^2$ at the Tevatron.

\section{Summary}

(1) The CKM unitarity relation $|V_{ud}|^2 + |V_{us}|^2 + |V_{ub}|^2 = 1$
seems valid, though questions remain.

\noindent
(2) Improved $V_{cd}$ and $V_{cs}$ values still are
consistent with $V_{cd} = - V_{us}$ and $V_{cs} = V_{ud}$.

\noindent
(3) Errors on $V_{cb} \sim 42 \times 10^{-3}$ from inclusive analyses
are being reduced; a $1 \sigma$ error of about 3\% is a conservative guess.
Lattice QCD will help exclusive analyses catch up.

\noindent
(4) Lower $V_{ub}$ values [$(3.26 \pm 0.62) \times 10^{-3}$] are obtained in
exclusive $b \to u$ decays than those [$(4.66 \pm 0.43) \times 10^{-3}$] in
inclusive decays.  One needs a better understanding of form factors and
quark-hadron duality.  An average $\sim 4.2 \times 10^{-3}$ is
known at present to $\simeq 15\%$.

\noindent
(5) Errors in $|V_{td}| \simeq (8.3^{+1.2}_{-1.8}) \times 10^{-3}$ (95\% c.l.)
are due mainly to lattice errors in $f_B \sqrt{B_B}$.  One expects a major
improvement from detection of $B_s$--$\overline{B}_s$ mixing.

\noindent
(6) The results of an SU(3) fit giving $\gamma = (63 \pm 6)^\circ$ are being
validated by the most recent measurements of rare $B$ decay branching ratios
\cite{Suprun}.  The detection of $b \to d$ penguins at expected rates
\cite{Sakai} has provided partial confirmation.

\noindent
(7) There is no evidence against $V_{ts} \simeq - V_{cb}$, $V_{tb} \simeq 1$.

\noindent
(8) It's time for a theory of quark masses and CKM elements!

\section*{Acknowledgments}

I thank C.-W. Chiang, M. Gronau, Z. Luo, D. Suprun for enjoyable collaborations
on some of the subjects mentioned here, A. Lenz and J. Prades for comments,
and Maury Tigner for
extending the hospitality of the Laboratory for Elementary-Particle Physics at
Cornell during part of this research.  This work was supported in part by the
U. S. Department of Energy under Grant No.\ DE-FG02-90ER40560 and in part by
the John Simon Guggenheim Memorial Foundation.

\end{document}